\documentclass[10pt]{iopart}
\usepackage{amssymb,graphicx,color,setspace}
\usepackage[colorlinks=true,urlcolor=blue,citecolor=blue]{hyperref}

\begin{document}

\title[Short-range PA from the inner wall of triplet $^{85}$Rb$_2$]{Short-range photoassociation from the inner wall of the lowest triplet potential of $^{85}$Rb$_2$}

\author{R. A. Carollo\footnote{Present Address: Department of Physics, Amherst College, PO Box 5000, Amherst, MA 01002-5000, USA}, J. L. Carini\footnote{Present Address: Pratt \& Whitney, 400 Main Street, East Hartford, CT 06118, USA}, E. E. Eyler, P. L. Gould, and W. C. Stwalley}
\address{Department of Physics, University of Connecticut, Storrs, CT 06269, USA}
\ead{carollo@phys.uconn.edu}

\date{\today}


\begin{abstract}
Ultracold photoassociation is typically performed at large internuclear separations, where the scattering wavefunction amplitude is large and Franck-Condon overlap is maximized. Recently, work by this group and others on alkali-metal diatomics has shown that photoassociation can efficiently form molecules at short internuclear distance in both homonuclear and heteronuclear dimers. We propose that this short-range photoassociation is due to excitation near the wavefunction amplitude maximum at the inner wall of the lowest triplet potential. We show that Franck-Condon factors from the highest-energy bound state can almost precisely reproduce Franck-Condon factors from a low-energy scattering state, and that both calculations match experimental data from the near-zero positive-energy scattering state with reasonable accuracy. We also show that the corresponding photoassociation from the inner wall of the ground-state singlet potential at much shorter internuclear distance is weaker and undetectable under our current experimental conditions. We predict from Franck-Condon factors that the strongest of these weaker short-range photoassociation transitions are one order of magnitude below our current sensitivity.
\end{abstract}

\noindent{\it Keywords\/}: photoassociation, ultracold molecules, transition matrix elements

\submitto{\jpb}
\ioptwocol

\maketitle

\section{Introduction}\label{shortrange:tradPA}

In recent work, our lab has performed a number of short-range photoassociation (PA) experiments in homonuclear $^{85}$Rb$_2$. In~\cite{bellos11} we studied blue-detuned and short-range photoassociation to the 4 different $\Omega = 0^+$, $0^-$, 1, and 2 components of the $1 \, ^3\Pi_g$ state, including various spectra. We also studied other short-range photoassociation spectra to the $2 \, ^1\Sigma_g^+$ state~\cite{bellos12,carollo13}, including a single rovibrational level in the short-range state ($v' = 111$, $J' = 5$) that is perturbed by a single, predominately long-range rovibrational level ($v' = 155$, $J' = 5$) of the $1 \, ^1 \Pi_{g, \, \Omega = 1}$ state, the two levels being separated by $\sim 0.017$ cm$^{-1}$.

When homonuclear photoassociation is done in the traditional long-range regime, the $\Sigma_n C_n/R^n$ form of the potential is different in the ground and excited states. For the lowest asymptote of two ground-state atoms, the long-range potential is due to London dispersion forces and takes the form
\begin{equation}
U = -\frac{C_6}{R^6} - \frac{C_8}{R^8} - \frac{C_{10}}{R^{10}} \, .
\end{equation}
These terms correspond to the dipole-dipole, dipole-quadrupole, and quadrupole-quadrupole plus dipole-octupole interactions. The excited state potentials, in addition to $C_6$, $C_8$, and $C_{10}$ terms, also contain a leading (positive or negative) $C_3/R^3$ term (for homonuclear molecules)~\cite{leroy09}, which dominates at long range.

This difference in functional form means that increasing detuning from the excited atomic asymptote reduces photoassociation efficiency. In heteronuclear molecules with no excited-state $C_3$ term~\cite{sadeghpour99}, efficiency still decreases due to the decreasing amplitude of the wavefunction in the scattering state at the internuclear distance of the excited-state outer turning point. In~\cite{pillet97}, Pillet \emph{et al.} derive an expression (equation (55)) for the homonuclear photoassociation efficiency and find that, in the limit of small $\Delta$, the efficiency is proportional to $\Delta^{-(4J'+7)/3}$, where $\Delta$ is the detuning below the atomic asymptote and $J'$ is the excited state rotational quantum number. This derivation assumes that PA can only occur below the asymptote, and implies that it is only strong enough to be observable near the atomic dissociation asymptote. Although true in many cases, both of these have been demonstrated to have exceptions in our work (in Rb$_2$~\cite{bellos11,bellos12}) and that of others (in LiCs~\cite{weidemuller08}, RbCs~\cite{gabbanini11,rbcs12,bruzewicz14}, and NaCs~\cite{bigelow11}). We note that the general approach in~\cite{pillet97} could be applied to short-range PA, but the expressions therein would need to be re-derived without restriction to long range.

This traditional view of photoassociation as predominantly a long-range process is based on several factors, including the number of atom pairs that exist at a given internuclear distance in the MOT and the fact that the amplitude of the nuclear wavefunction is small at short range and much larger at long range. However, there are also several mechanisms that provide exceptions to this assumption. One is that there is a significant local maximum in the square of the wavefunction amplitude at the inner wall of the $a \, ^3 \Sigma_u^+$ potential in $^{85}$Rb$_2$, roughly at $9.7 \, a_0$, as can be seen in Figure~\ref{fcfmodel}. This can enhance the PA rate if the excited state has significant short-range amplitude as well, such as exists behind a short- or intermediate-range potential barrier. As was discussed in~\cite{bellos12}, there is also a $g$-wave shape resonance in the triplet potential because of its centrifugal barrier near $80 \, a_0$, which can provide a second source of short-range PA enhancement. Similar considerations apply for the zero-energy wavefunction in the $a \, ^3\Sigma^+$ state of many other homo- and heteronuclear alkali dimers. A third mechanism for enhancing PA to short-range states is excited-state resonant coupling, which can combine efficient long-range excitation with efficient short-range decay in both homonuclear~\cite{carollo13,pechkis07} and heteronuclear~\cite{bigelow11,banerjee12,stwalley10} cases. A fourth path to efficient formation of short-range states is Feshbach-optimized PA (FOPA)~\cite{pellegrini08,abraham15}.

Here we explore the first of these mechanisms, photoassociation from the inner turning point of the continuum wavefunction. To estimate relative PA rates to various excited-state vibrational levels, we take advantage of the continuity of the absorption (or emission) cross section as one passes through a dissociation limit, which was pointed out by Allison and Dalgarno~\cite{dalgarno71} and Smith~\cite{smith71} some 45 years ago, using as examples the H$_2$ (a ``VUV'' alkali metal diatomic) Lyman bands and O$_2$ Schumann-Runge bands.  While~\cite{dalgarno71} used a Morse potential for the excited B state, which is unphysical at long range, the continuity behavior for the $B$ state of H$_2$ with correct long-range behavior was published two years later~\cite{stwalley73}, and quantitatively showed continuity at dissociation.  In particular, we take advantage of the fact that the short-to-intermediate range behavior of the wavefunction of the highest long-range bound level (readily calculated by standard programs such as LEVEL 8.2~\cite{leroylevel82}) is very nearly identical to the short-to-intermediate range behavior of the continuum wavefunction at near-zero-energy, i.e. ultracold energies, as shown in Figure~\ref{fcfmodel}. Note that for most cases of photoassociation at long range, where the last bound level and near-zero-energy continuum wavefunctions differ, the two Franck-Condon factor (FCF) calculations will no longer agree quantitatively.

A direct consequence is that the ultracold limit of the FCF for free $\rightarrow$ bound photoassociation occurring at relatively short range is, within a scaling factor, well approximated by the bound $\rightarrow$ bound FCF for a corresponding transition from the uppermost vibrational level, $v''_{max}$:
\begin{equation}\label{fcfapprox}
\left| \left< k'', J'' = 0| v', J' = 1 \right> \right|^2 \propto \left| \left< v''_{max}, J'' = 0| v', J' = 1 \right> \right|^2
\end{equation}
where the radial wavefunctions for $\left| k'', J'' = 0 \right>$ (free) and $\left| v''_{max}, J'' = 0 \right>$ (bound) are very similar, as shown in Figure~\ref{fcfmodel}, as are the FCFs listed in Equation~\ref{fcfapprox}. Here the ultracold limit is taken to be $\sim 120 \, \mu$K, the approximate temperature of our magneto-optical trap.

\section{Photoassociation Model}\label{shortrange:model}

\begin{figure}[tb]
\begin{center}
\includegraphics[width=\columnwidth]{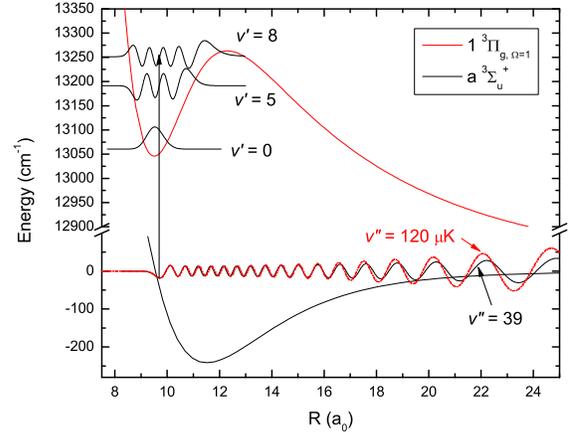}
\caption{(Color online) Wavefunctions of the $a \, ^3\Sigma_u^+$, $v'' = 39$ vibrational level of $^{85}$Rb$_2$ and the 120 $\mu$K scattering state, each with zero rotational/collisional angular momentum, and three rovibrational levels of the $1 \, ^3\Pi_{g, \, \Omega = 1}$ state, the $v' = 0$, 5, and 8 levels, each with $J' = 1$ are shown. The $v'' = 39$ wavefunction closely resembles a low-energy continuum wavefunction throughout the range of R shown here. The potentials shown are from~\cite{tiemann10} for the $a \, ^3\Sigma_u^+$ state and from~\cite{bellos11} (and [17] therein) for the $1 \, ^3\Pi_{g, \, \Omega = 1}$ state. Note that the $v' = 0$ and 5 wavefunctions are similar to harmonic oscillator wavefunctions, while $v' = 8$, near the top of a potential barrier, is more asymmetric toward larger distance. For $v' = 0$ (and 2, 4, and even 6), there is a significant FCF from $v'' = 39$, while it is small for $v' = 5$ (and 3 and 1, as shown in Figure~\ref{fcfdata}). Here, the 120 $\mu$K wavefunction is scaled to match the $v'' = 39$ wavefunction at short $R$ to demonstrate their similarity. For all other uses, this state was scaled as described in the text. It should be noted that the $v'' = 39$ wavefunction amplitude increases dramatically at longer range out to its maximum at $\sim 62$ a$_0$.}
\label{fcfmodel}
\end{center}
\end{figure}

As stated in Section~\ref{shortrange:tradPA}, we often operate under circumstances where traditional assumptions about photoassociation do not apply because of a local short-range maximum in the target-state wavefunction. We thus introduce a simple model to predict the efficiency of photoassociation in the short-range or blue-detuned regions that were formerly considered inaccessible. Our model assumes that the inner turning point  of the near-zero-energy free wavefunction, or incoming scattering state, creates a population of closely-spaced atom pairs sufficient to form molecules at relatively short range. If the excited state of interest has significant amplitude at short range, the PA rate can be enhanced. Further, for sufficiently small energy differences, the wavefunction at short range is equivalent both above and below the atomic dissociation asymptote, allowing the highest bound state to be used for FCF calculations. An example of short-range excitation to the $^{85}$Rb$_2$ $1 \, ^3\Pi_{g, \, \Omega = 1}$ state, which lies above the $5s_{1/2} + 5p_{3/2}$ asymptote, is shown in Figure~\ref{fcfmodel}.  We believe that this model is an additional new and useful `shortcut' for understanding rovibronic spectroscopy of ultracold molecules~\cite{stwalley12} and enhancing their formation.

In a time-dependent view, this can be interpreted as meaning that atom pairs spend significant time in close proximity when they collide at short range. This period of slow movement at short range means that there is always a non-negligible fraction of atoms available to interact at these internuclear distances. Since the lowest triplet potential inner wall is at significantly longer range than the inner wall of the singlet $X$ state, colliding triplet pairs of ultracold atoms are able to access many states of interest that are inaccessible to ultracold colliding pairs of singlet character. By contrast, singlet wavefunctions have rapid, low-amplitude oscillations at most internuclear distances at which excited states are present, drastically reducing the transition probability. A partial exception is in the region of an intermediate-range barrier maximum in the $B \, ^1\Pi_u$ state in $^{85}$Rb$_2$, discussed in Section~\ref{shortrange:comparison}.

Thus, any PA transitions that are to be studied at short range must be accessible from the lowest triplet potential. In particular, this means that the transition must be allowed from the lowest triplet state in the appropriate Hund's case. For pairs of identical ground-state alkali atoms, this implies PA to a triplet \emph{gerade} state in Hund's case (a), or a $0_g^+$, $0_g^-$, $1_g$, or $2_g$ state in Hund's case (c). Although for heavy alkalis such as $^{85}$Rb the singlet $\nleftrightarrow$ triplet selection rule is weakened (or equivalently, case (a) quantum numbers are no longer perfectly ``good'' and case (c) quantum numbers may be more appropriate), the $g \leftrightarrow u$ selection rule is still strong in a homonuclear system.

A simple way to use this model to calculate relative excitation probabilities from free triplet atoms to a given excited state is to calculate the FCF between the highest bound level of the lowest triplet state and each level of the target excited state. It is slightly more accurate to calculate the square of the overlap integral from the true near-zero-energy (free) state, but it is more convenient to use a bound $\rightarrow$ bound calculation program such as LEVEL 8.2~\cite{leroylevel82}. With appropriate scaling, the highest bound vibrational level is nearly identical at short range to the near-zero-energy scattering state, as shown in Figure~\ref{fcfmodel}, due to the steepness of the repulsive wall and the similarity in their energies. At our experimental temperature of $\sim 120 \, \mu$K, the thermal population distribution has a peak at 2.5 MHz, while the $v'' = 39$ level is bound by 0.007238 cm$^{-1}$, or 217 MHz, for a difference of less than $\sim 220$ MHz. The scattering state has been calculated using the code developed in~\cite{kosloff06,carini15a,carini15b}. Franck-Condon factors were calculated from this box-normalized free wavefunction and were scaled to give the same FCF at $v' = 2$ as produced by the bound state calculation. We show a comparison of such an FCF calculation to experimental data in Section~\ref{shortrange:comparison}.

As with any use of a Franck-Condon factor, the $R$-dependence of the transition dipole moment may play an important role that is neglected by our approximation. The $2 \, ^1 \Sigma_g^+ \sim 2 \, (0_g^+)$ state studied in~\cite{bellos12,carollo13} is an excellent example. The case (c) transition dipole is constant and large at long range, but at short range (in the region of our experiment) it drops rapidly toward zero~\cite{allouche12}. This caused a drop-off in our signal for shorter-range and more deeply-bound vibrational levels, and reflects a transition from the allowed transition in case (c) to a region where case (a) better represents the coupling and the transition is singlet $\leftrightarrow$ triplet forbidden.

The calculation of FCFs between the $v'' = 39$ level of the $a \, ^3\Sigma_u^+$ state and the $v'$ levels of the $1 \, ^3 \Pi_{g, \, \Omega = 1}$ state is based on the experimentally determined potential of the $a \, ^3\Sigma_u^+$ state~\cite{tiemann10} and an \emph{ab initio} potential of the $1 \, ^3 \Pi_{g, \, \Omega = 1}$ state calculated by Dulieu and Gerdes that reproduces the experimental vibrational and rotational constants fairly accurately (see Table 1 of~\cite{bellos11}). The $a \, ^3\Sigma_u^+$ state potential produces a scattering length of $a_T = -371$ $a_0$ in our calculations, which agrees well with published scattering lengths for $^{85}$Rb$_2$ in Roberts \emph{et al.} ($a_T = -369 \pm 16$ $a_0$~\cite{weiman98}) and van Kempen \emph{et al.} ($a_T = -388 \pm 3$ $a_0$~\cite{verhaar02}). The electronic transition dipole moment is generally not expected to vary significantly in the small region of overlap of vibrational wavefunctions $\psi'(R)$ and $\psi''(R)$. In this case, the transition dipole moments for $1 \, ^3 \Pi_g \leftarrow a \, ^3\Sigma_u^+$ (and $B \, ^1\Pi_u \leftarrow X \, ^1\Sigma_g^+$, discussed below) are quite similar, $\sim 3.5$ a.u. and $\sim 4$ a.u., respectively, at the internuclear distance of the transition~\cite{allouche12}.

The potential energy curve used here and plotted in our Figure~\ref{fcfmodel} for the $a \, ^3\Sigma_u^+$ state from~\cite{tiemann10} is a high-quality experimentally-based potential (note that Figure 1 of~\cite{tiemann10} does not have the correct R axis values; for example, the Table II value of $R_e  = 5.07$ \AA $\;= 11.5160 \, a_0$ is correct in our Figure~\ref{fcfmodel}, but is $\sim10 \, a_0$ in Figure 1 of~\cite{tiemann10}).  In particular,~\cite{tiemann10} argues that their results, compared to earlier results, ``are significantly improved close to the atomic asymptote by including data on the mixed singlet-triplet levels of this study and data on Feshbach resonances from various other sources.''  This near dissociation region is exactly the region of greatest significance for our present study.

\section{Comparison to Data}\label{shortrange:comparison}

To test whether these FCFs can predict the relative efficiency of short-range PA to various levels, we looked at several lines that our group previously detected~\cite{bellos11}. The target state is the $\Omega = 1$ component of the $1 \, ^3 \Pi_g$ manifold of $^{85}$Rb$_2$. This state is blue-detuned above the $5s + 5s_{3/2}$ asymptote.

Our experiment was carried out under conditions similar to the original work, and a detailed description can be found in~\cite{bellos11,bellos12}. The molecules were formed in a $^{85}$Rb MOT of typically $8 \times 10^7$ atoms and a density of $1 \times 10^{11}$ cm$^{-3}$ at $\sim 120 \, \mu$K. The excitation was performed with a fiber-coupled photoassociation laser (Coherent 899-29 Ti:Sapphire) delivering 650 mW to the experimental chamber. After photoassociation, the molecules rapidly decay to deeply-bound levels of the $a \, ^3\Sigma_u^+$ state and are detected via pulsed ionization. Ions are detected on a discrete-dynode multiplier (ETP model 14150) and spectra are acquired via a boxcar integrator which gates the molecular time-of-flight signal.

To ensure accurate relative line height measurements, detection was done using photoionization with a pulsed 355 nm UV tripled Nd:YAG laser at 3.6 mJ/pulse. This photon energy corresponds to $\sim 28,169$ cm$^{-1}$. Based on the data we reported in~\cite{bellos13}, the $v = 0$ level of the Rb$_2^+$ ground-state potential is no higher than 27,383.2 cm$^{-1}$. Accounting for the $\sim 234.7$ cm$^{-1}$ binding energy of the $a \, ^3\Sigma_u^+$, $v'' = 0$ level, up to 27,617.9 cm$^{-1}$ could be necessary to ionize. Our UV photon energy is $\sim 551$ cm$^{-1}$ above this, and thus all ionization should be single-photon and line strengths should be independent of any intermediate-state resonances. Measured line strengths should therefore reflect the true relative transition probabilities. We note that the PA rates for these lines are too small to be observed via trap loss.

\begin{figure}[tb]
\begin{center}
\includegraphics[width=\columnwidth]{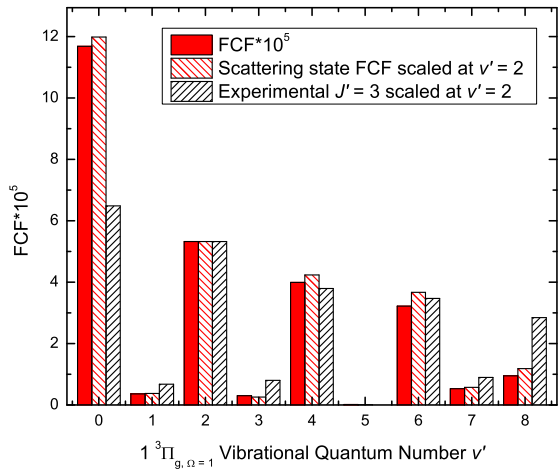}
\caption{(Color online) Comparison of FCF calculations to experimental data. For each vibrational level in the excited state, three bars are shown. The first bar (solid red) shows FCFs to the $1 \, ^3 \Pi_g, \, \Omega = 1$ state from the highest bound state of the lowest triplet potential, $a \, ^3 \Sigma _u^+, \, v''=39$, which closely approximates the zero-energy scattering state. The excited state potentials are from~\cite{bellos11}. The second bar (hashed red) is calculated from the 120 $\mu$K scattering state, and scaled to match the bound-state calculation at $v' = 2$. The experimental data, shown in the third (hashed black) bar, are obtained using PA from a Ti:Sapphire laser detected by photoionization at 355 nm from a tripled Nd:YAG laser, and are also normalized to agree with the bound-state FCFs at $v' = 2$. The FCFs are calculated for $J'' = 0$ and $J' = 1$.}
\label{fcfdata}
\end{center}
\end{figure}

Each vibrational level was measured and its line height at $J' = 3$ recorded. As discussed in Section~\ref{shortrange:model}, FCFs were calculated for the same transitions by using the $a \, ^3\Sigma_u^+$, $v'' = 39$ level as a proxy for the near-zero-energy scattering state. The FCFs from the 120 $\mu$K scattering state itself were also calculated. A comparison of these data and the FCF calculations is shown in Figure~\ref{fcfdata}. The experimental data are scaled to match the bound-state FCF calculations at $v' = 2$ to better show the quality of the comparison. The scattering-state FCFs are also scaled to the bound-state FCFs to ensure comparable normalization. It should be noted that levels with FCFs as low as $2.99 \times 10^{-6}$ ($v' = 3$) are detected. However, $v' = 5$, with a FCF of $1.70 \times 10^{-7}$, is not detected.

The prominent alternation in FCF seen in Figure~\ref{fcfdata} bears closer examination. The strong barrier in the $1 \, ^3 \Pi_{g, \, \Omega = 1}$ state creates a short-range well that differs little from a harmonic potential, and the resulting wavefunctions also closely approximate those of a harmonic oscillator. In this case, the strong maximum of the $v' = 0$ wavefunction is replaced by a node in $v' = 1$, with two weaker and opposite-signed extrema adjacent. If the FCF from the initial state to $v' = 0$ is coming from a localized region of the $v'' = 39$ wavefunction, the expected result is a sharp reduction (nearly cancellation) in the FCF to $v' = 1$. Similarly, at $v' = 2$ a new, albeit weaker, extremum will occupy the position of the $v' = 0$ maximum, which gives rise to an alternating series of FCFs with gradually decaying contrast. The strong alternation actually observed thus supports the hypothesis that the PA in this experiment is due to the local maximum at the inner turning point of the $a \, ^3\Sigma_u^+$ state.

As further evidence in support of our model, we also scanned the predicted locations of quasibound vibrational levels of the $B \, ^1\Pi_u$ state, which corresponds to the Hund's case (c) $3 \, (1_u)$ state. Much like the $1 \, ^3 \Pi_g$ state, the $B$ state has an intermediate-range barrier (17.3 $a_0$~\cite{amiot97}) and is repulsive at long range, causing vibrational states to be confined to short range. Levels of this state were accurately measured by Amiot and Verg{\`e}s using optical-optical double resonance and Fourier-transform spectroscopy~\cite{amiot97,amiot90}. If our model is accurate, however, any photoassociation to this state must, by selection rules, originate from free atoms of \emph{gerade} symmetry, and therefore must come from the $X \, ^1\Sigma_g^+$ state. This state has a very short-range inner wall, and does not give highly-enhanced FCFs for excitation to higher bound states. The relatively larger FCFs high in the $B$ state come predominantly from the outer turning points at the potential barrier, where the $X$ state wavefunction is beginning to grow in amplitude.

\begin{figure}[tb]
\begin{center}
\includegraphics[width=\columnwidth]{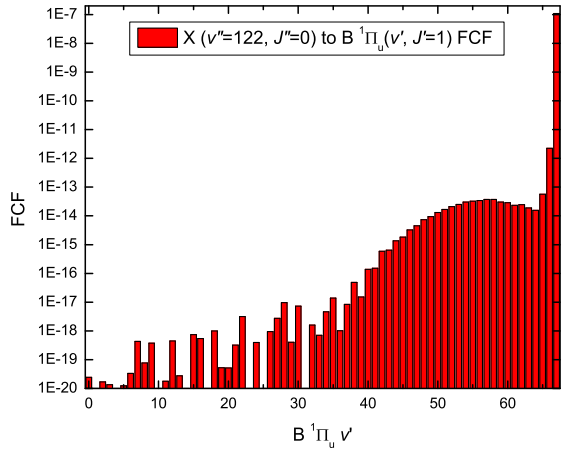}
\caption{(Color online) FCFs for excitation to the $B \, ^1\Pi_u$ state from the $X \, ^1\Sigma_g^+$ state. Although the last two levels have dramatically increasing FCFs, it should be noted that, since they are quasibound, this transition strength is spread over a larger energy range. Widths of low-$J$ levels were not experimentally observed, although some high-$J$ widths are reported~\cite{amiot97}.}
\label{BfromX}
\end{center}
\end{figure}

The FCFs for transitions to $J' = 1$ levels of the $B \, ^1\Pi_u$ state from $v'' = 122$, $J'' = 0$ of the $X \, ^1\Sigma_g^+$ state are shown in Figure~\ref{BfromX}. The potential used to calculate these FCFs was obtained from the ``inverted perturbation approach'' (IPA) potential in~\cite{amiot97}, with the top of the barrier and repulsive wall from the \emph{ab initio} potential in~\cite{bellos11}. The \emph{ab initio} barrier was translated to match the energy and slope of the IPA potential, and the vibrational eigenstates were shown to match experimental values closely. No attempt was made to reproduce the linewidths observed in high rotational levels of $v' = 66$ and 67. Even the largest FCF to this state is $\sim 1 \times 10^{-7}$, and the vast majority of FCFs are $< 1 \times 10^{-13}$. These are well below the FCFs of levels we are able to detect based on the $1 \, ^3 \Pi_g$ state data above. FCFs can be directly compared to make this determination because the transition dipole moments are so similar. We discuss our detection sensitivity below.

To briefly summarize, the $B$ state has a short-range potential well that is amenable to our short-range PA efficiency calculation technique. Due to symmetry considerations, only atom pairs of singlet-\emph{gerade} character may be excited to this state. As the $X$ state's near-zero-energy inner turning point is at shorter range than the classically allowed region, very poor PA efficiency is predicted. In fact, the dominant contribution to the largest FCFs for the $B$ state come from the outer turning points, as in traditional long-range PA, although these turning points are at the intermediate-range potential barrier.

We have measured the statistics of our laser scan data in the vicinity of the $v' = 65$ level of the $B$ state. Our scans have noise with an average standard deviation of 0.18 ions per shot across several scans. This is easily capable of $5 \sigma$ detection of a single ion per laser shot, and several stray ions are indeed observed within the correct boxcar gate during the scan. These features are not repeatable, and thus cannot be spectroscopic features. Additionally, they show an exponential drop-off characteristic of our real-time boxcar shot averaging rate (typically a 10-shot rolling exponential average). Other than these, no lines are detected near the predicted $v' = 65$ energy.

In a similar scan, the $1 \, ^3 \Pi_{g, \, \Omega = 1}$, $v' = 8$ level is detected. Between lines, the scan background shows a standard deviation of 0.33 ions per shot, comparable to the average mentioned above. The peak line height is 9.18 ions above the baseline. Based on the FCF of $9.48 \times 10^{-6}$ for this level, we are sensitive to FCFs as small as $9.3 \times 10^{-7}$ with average noise levels. Thus a non-detection of $B$ state levels is consistent with our FCF calculation.

\section{Conclusion}

We have presented a model of short-range photoassociation in an alkali dimer that is both conceptually and computationally simple. It applies when the square of the excited-state target wavefunction has an appreciable local maximum near the internuclear distance of the scattering state's short-range turning point. In this model, short-range excitation is proportional to the FCF of a transition from the near-zero-energy continuum of the lowest triplet state (or the highest bound vibrational level, which we have shown to produce nearly-identical results), so long as the transition is allowed in the appropriate Hund's case. Generally, the correct coupling description is Hund's case (a) or (c), with case (c) likely being more important in heavier alkalis with strong spin-orbit coupling.

We have presented experimental data in $^{85}$Rb$_2$ using detection via single-photon ionization that should accurately reflect molecule production regardless of vibrational level. The FCF predictions from our model show quite reasonable agreement with the observed $1 \, ^3 \Pi_{g, \, \Omega = 1}$ state lines, and correctly predict that lines from the $B \, ^1\Pi_u$ state should not be observed.

\ack
The concepts presented here were initially stimulated by~\cite{dalgarno71} and decades of conversations of William C. Stwalley with Arthur Allison and Alex Dalgarno. This research was funded with support from the National Science Foundation grant numbers PHY-1208317 and PHY-1506244 and Air Force Office of Scientific Research grant number FA9550-09-1-0588.

\section*{References}

\bibliographystyle{iopart-num}
\bibliography{ultracold_references}

\providecommand{\newblock}{}
\begin{thebibliography}{10}
\expandafter\ifx\csname url\endcsname\relax
  \def\url#1{{\tt #1}}\fi
\expandafter\ifx\csname urlprefix\endcsname\relax\def\urlprefix{URL }\fi
\providecommand{\eprint}[2][]{\url{#2}}

\bibitem{bellos11}
Bellos M~A, Rahmlow D, Carollo R, Banerjee J, Dulieu O, Gerdes A, Eyler E~E,
  Gould P~L and Stwalley W~C 2011 {\em Phys. Chem. Chem. Phys.\/} {\bf 13}(42)
  18880--18886 \urlprefix\url{http://dx.doi.org/10.1039/C1CP21383K}

\bibitem{bellos12}
Bellos M~A, Carollo R, Rahmlow D, Banerjee J, Eyler E~E, Gould P~L and Stwalley
  W~C 2012 {\em Phys. Rev. A\/} {\bf 86}(3) 033407
  \urlprefix\url{http://link.aps.org/doi/10.1103/PhysRevA.86.033407}

\bibitem{carollo13}
Carollo R, Bellos M~A, Rahmlow D, Banerjee J, Eyler E~E, Gould P~L and Stwalley
  W~C 2013 {\em Phys. Rev. A\/} {\bf 87}(2) 022505
  \urlprefix\url{http://link.aps.org/doi/10.1103/PhysRevA.87.022505}

\bibitem{leroy09}
Le~Roy R~J, Dattani N~S, Coxon J~A, Ross A~J, Crozet P and Linton C 2009 {\em
  The Journal of Chemical Physics\/} {\bf 131} 204309
  \urlprefix\url{http://scitation.aip.org/content/aip/journal/jcp/131/20/10.1063/1.3264688}

\bibitem{sadeghpour99}
Marinescu M and Sadeghpour H~R 1999 {\em Phys. Rev. A\/} {\bf 59}(1) 390--404
  \urlprefix\url{http://link.aps.org/doi/10.1103/PhysRevA.59.390}

\bibitem{pillet97}
Pillet P, Crubellier A, Bleton A, Dulieu O, Nosbaum P, Mourachko I and
  Masnou-Seeuws F 1997 {\em Journal of Physics B: Atomic, Molecular and Optical
  Physics\/} {\bf 30} 2801
  \urlprefix\url{http://stacks.iop.org/0953-4075/30/i=12/a=010}

\bibitem{weidemuller08}
Deiglmayr J, Grochola A, Repp M, M{\"o}rtlbauer K, Gl{\"u}ck C, Lange J, Dulieu
  O, Wester R and Weidem{\"u}ller M 2008 {\em Phys. Rev. Lett.\/} {\bf 101}(13)
  133004 \urlprefix\url{http://link.aps.org/doi/10.1103/PhysRevLett.101.133004}

\bibitem{gabbanini11}
Gabbanini C and Dulieu O 2011 {\em Phys. Chem. Chem. Phys.\/} {\bf 13}(42)
  18905--18909 \urlprefix\url{http://dx.doi.org/10.1039/C1CP21497G}

\bibitem{rbcs12}
Ji Z, Zhang H, Wu J, Yuan J, Yang Y, Zhao Y, Ma J, Wang L, Xiao L and Jia S
  2012 {\em Phys. Rev. A\/} {\bf 85}(1) 013401
  \urlprefix\url{http://link.aps.org/doi/10.1103/PhysRevA.85.013401}

\bibitem{bruzewicz14}
Bruzewicz C~D, Gustavsson M, Shimasaki T and DeMille D 2014 {\em New Journal of
  Physics\/} {\bf 16} 023018
  \urlprefix\url{http://stacks.iop.org/1367-2630/16/i=2/a=023018}

\bibitem{bigelow11}
Zabawa P, Wakim A, Haruza M and Bigelow N~P 2011 {\em Phys. Rev. A\/} {\bf
  84}(6) 061401
  \urlprefix\url{http://link.aps.org/doi/10.1103/PhysRevA.84.061401}

\bibitem{pechkis07}
Pechkis H~K, Wang D, Huang Y, Eyler E~E, Gould P~L, , Stwalley W~C and Koch C~P
  2007 {\em Phys. Rev. A\/} {\bf 76} 022504
  \urlprefix\url{http://link.aps.org/doi/10.1103/PhysRevA.76.022504}

\bibitem{banerjee12}
Banerjee J, Rahmlow D, Carollo R, Bellos M, Eyler E~E, Gould P~L and Stwalley
  W~C 2012 {\em Phys. Rev. A\/} {\bf 86}(5) 053428
  \urlprefix\url{http://link.aps.org/doi/10.1103/PhysRevA.86.053428}

\bibitem{stwalley10}
Stwalley W~C, Banerjee J, Bellos M, Carollo R, Recore M and Mastroianni M 2010
  {\em J. Phys. Chem. A\/} {\bf 114} 81--86
  \urlprefix\url{http://pubs.acs.org/doi/abs/10.1021/jp901803f}

\bibitem{pellegrini08}
Pellegrini P, Gacesa M and C{\^o}t{\'e} R 2008 {\em Phys. Rev. Lett.\/} {\bf
  101}(5) 053201
  \urlprefix\url{http://link.aps.org/doi/10.1103/PhysRevLett.101.053201}

\bibitem{abraham15}
Krzyzewski S~P, Akin T~G, Dizikes J, Morrison M~A and Abraham E~R~I 2015 {\em
  Phys. Rev. A\/} {\bf 92}(6) 062714
  \urlprefix\url{http://link.aps.org/doi/10.1103/PhysRevA.92.062714}

\bibitem{dalgarno71}
Allison A~C and Dalgarno A 1971 {\em The Journal of Chemical Physics\/} {\bf
  55} 4342--4344
  \urlprefix\url{http://scitation.aip.org/content/aip/journal/jcp/55/9/10.1063/1.1676757}

\bibitem{smith71}
Smith A~L 1971 {\em The Journal of Chemical Physics\/} {\bf 55} 4344--4350
  \urlprefix\url{http://scitation.aip.org/content/aip/journal/jcp/55/9/10.1063/1.1676758}

\bibitem{stwalley73}
Allison A~C and Stwalley W~C 1973 {\em The Journal of Chemical Physics\/} {\bf
  58} 5187--5188
  \urlprefix\url{http://scitation.aip.org/content/aip/journal/jcp/58/11/10.1063/1.1679123}

\bibitem{leroylevel82}
Le{ }Roy R~J 2014 {\em LEVEL 8.2: A Computer Program for Solving the Radial
  Schr{\"o}dinger Equation for Bound and Quasibound Levels\/} University of
  Waterloo Chemical Physics Research Report {CP}-668 see
  http://scienide2.uwaterloo.ca/\%7erleroy/level/
  \urlprefix\url{http://scienide2.uwaterloo.ca/\%7erleroy/level/}

\bibitem{stwalley12}
Stwalley W~C, Bellos M, Carollo R, Banerjee J and Bermudez M 2012 {\em
  Molecular Physics\/} {\bf 110} 1739--1755
  \urlprefix\url{http://www.tandfonline.com/doi/abs/10.1080/00268976.2012.676680}

\bibitem{tiemann10}
Strauss C, Takekoshi T, Lang F, Winkler K, Grimm R, Hecker~Denschlag J and
  Tiemann E 2010 {\em Phys. Rev. A\/} {\bf 82}(5) 052514
  \urlprefix\url{http://link.aps.org/doi/10.1103/PhysRevA.82.052514}

\bibitem{kosloff06}
Kallush S and Kosloff R 2006 {\em Chemical Physics Letters\/} {\bf 433} 221 --
  227 ISSN 0009-2614
  \urlprefix\url{http://www.sciencedirect.com/science/article/pii/S0009261406016940}

\bibitem{carini15a}
Carini J~L, Kallush S, Kosloff R and Gould P~L 2015 {\em Phys. Rev. Lett.\/}
  {\bf 115}(17) 173003
  \urlprefix\url{http://link.aps.org/doi/10.1103/PhysRevLett.115.173003}

\bibitem{carini15b}
Carini J~L, Kallush S, Kosloff R and Gould P~L 2016 {\em The Journal of
  Physical Chemistry A\/} {\bf 120} 3032--3041
  \urlprefix\url{http://dx.doi.org/10.1021/acs.jpca.5b10088}

\bibitem{allouche12}
Allouche A~R and Aubert-Fr\'{e}con M 2012 {\em The Journal of Chemical
  Physics\/} {\bf 136} 114302
  \urlprefix\url{http://scitation.aip.org/content/aip/journal/jcp/136/11/10.1063/1.3694014}

\bibitem{weiman98}
Roberts J~L, Claussen N~R, Burke J~P, Greene C~H, Cornell E~A and Wieman C~E
  1998 {\em Phys. Rev. Lett.\/} {\bf 81}(23) 5109--5112
  \urlprefix\url{http://link.aps.org/doi/10.1103/PhysRevLett.81.5109}

\bibitem{verhaar02}
van Kempen E~G~M, Kokkelmans S~J~J~M~F, Heinzen D~J and Verhaar B~J 2002 {\em
  Phys. Rev. Lett.\/} {\bf 88}(9) 093201
  \urlprefix\url{http://link.aps.org/doi/10.1103/PhysRevLett.88.093201}

\bibitem{bellos13}
Bellos M~A, Carollo R, Banerjee J, Ascoli M, Allouche A~R, Eyler E~E, Gould P~L
  and Stwalley W~C 2013 {\em Phys. Rev. A\/} {\bf 87}(1) 012508
  \urlprefix\url{http://link.aps.org/doi/10.1103/PhysRevA.87.012508}

\bibitem{amiot97}
Amiot C and Verg{\`e}s J 1997 {\em Chemical Physics Letters\/} {\bf 274} 91 --
  98 ISSN 0009-2614
  \urlprefix\url{http://www.sciencedirect.com/science/article/pii/S0009261497006349}

\bibitem{amiot90}
Amiot C 1990 {\em The Journal of Chemical Physics\/} {\bf 93} 8591--8604
  \urlprefix\url{http://scitation.aip.org/content/aip/journal/jcp/93/12/10.1063/1.459246}

\end{thebibliography}

\end{document}